\documentclass[aps,prd,twocolumn,showpacs,preprintnumbers,amsmath,amssymb]{revtex4}

\usepackage{amsmath,amssymb}

\usepackage{color}
\usepackage[dvips]{graphicx}
\usepackage{subfigure}

\newcommand{\kX}{\mbox{$\vec{k}_{\rm X}$}}
\newcommand{\lsim}{\mathrel{\mathop{\kern 0pt \rlap
  {\raise.2ex\hbox{$<$}}}
  \lower.9ex\hbox{\kern-.190em $\sim$}}}
\newcommand{\gsim}{\mathrel{\mathop{\kern 0pt \rlap
  {\raise.2ex\hbox{$>$}}}
  \lower.9ex\hbox{\kern-.190em $\sim$}}}
\newcommand{\sigmav}{\langle \sigma_{\rm ann} v \rangle}
\newcommand{\xisigmav}{\xi^2\langle \sigma_{\rm ann} v \rangle}

\newcommand{\pbar}{\bar{p}}

\newcommand{\beq}{\begin{equation}}
\newcommand{\eeq}{\end{equation}}
\newcommand{\bea}{\begin{eqnarray}}
\newcommand{\ena}{\end{eqnarray}}

\def\pbar{$\overline{p}$}
\def\nbar{$\overline{n}$}

\def\dbar{$\overline{d}$}
\begin{document}

\preprint{DFTT 06/2007}

\title{Antideuteron fluxes from dark matter annihilation in diffusion models}

\author{F. Donato}
\email{donato@to.infn.it}
\affiliation{Dipartimento di Fisica Teorica, Universit\`a di Torino \\
Istituto Nazionale di Fisica Nucleare, via P. Giuria 1, I--10125 Torino, Italy}

\author{N. Fornengo}
\email{fornengo@to.infn.it}
\affiliation{ Dipartimento di Fisica Teorica, Universit\`a di Torino \\
Istituto Nazionale di Fisica Nucleare, via P. Giuria 1, I--10125 Torino, Italy}

\author{D. Maurin}
\email{dmaurin@lpnhe.in2p3.fr}
\affiliation{ Laboratoire de Physique Nucl\'eaire et Hautes Energies,
CNRS-IN2P3/Universit\'e Paris VI et VII, 4 Place Jussieu, Tour 33, 75252 Paris Cedex
05, France}

\date{\today}

\begin{abstract} 
Antideuterons are among the {most} promising galactic cosmic ray-related targets 
for dark matter indirect detection. {Currently} only upper limits exist on {the} flux,
but the development of new experiments, such as GAPS and AMS--02,  
provides exciting perspectives for a positive measurement in the near future.
{In this Paper, we present a novel and updated calculation of
both the secondary and primary \dbar\ fluxes. We employ
a two--zone diffusion model which successfully reproduces cosmic--ray nuclear
data and the observed antiproton flux}. We review the nuclear and astrophysical
uncertainties and provide an up to date secondary ({\em i.e.} background) antideuteron flux. 
{The primary ({\em i.e.} signal) contribution is calculated for generic WIMPs 
annihilating in the galactic halo: we explicitly consider and quantify the various sources of
uncertainty in the theoretical evaluations.
Propagation uncertainties, as is the case of
antiprotons, are sizeable. Nevertheless, antideuterons offer an exciting target for indirect dark
matter detection for low and intermediate mass WIMP dark matter. We then show
the reaching capabilities of the future experiments for neutralino dark matter 
in a variety of supersymmetric models.}
\end{abstract}


\maketitle


\section{Introduction}
\label{sect:intro}
The  identification and the understanding of the nature of dark matter (DM) 
is one of the deepest open problems, together with the solution to 
the dark energy mystery, in the fundamental physics research.
Many experimental efforts {devoted} to the detection of the astronomical 
dark matter in the halo of our and  nearby galaxies have been carried out
in underground laboratories, in large-area surface telescopes as well as in 
space. In the near future, the LHC will provide us with invaluable 
information on particle physics extending beyond the Standard Model, 
thus probing a wide class of theoretical models hosting the most viable DM 
candidates. 

The indirect dark matter detection is based on the search for anomalous
components due to the annihilation of DM pairs in the galactic halo, 
in addition to the standard astrophysical production of neutrinos, 
gamma rays and light antimatter in cosmic rays. 
Data on neutrinos, gamma rays, positrons and antiprotons are already available 
at a sensitivity level allowing some inspection on possible exotic contributions. 
In the seminal Paper \cite{2000PhRvD..62d3003D}, it was proposed   to look for
cosmic antideuterons  (\dbar) as a possible indirect signature for galactic  dark matter.
It was shown that the  antideuteron spectra deriving from DM annihilation  is
expected to be much flatter than the  standard astrophysical component at low
kinetic energies, $T_{\overline{d}}\lsim$ 2-3 GeV/n. 
This argument motivated the proposal of a new space-borne experiment
\cite{2002ApJ...566..604M,2004NIMPB.214..122H,2006JCAP...01..007H} looking for
cosmic antimatter (antiproton and antideuteron) and having the potential to
discriminate between standard and exotic components for a wide range of DM
models. {AMS--02 has also interesting capabilities of looking for cosmic antideuterons 
\citep{2007arXiv0710.0993A}}.
Antideuterons have not been measured so far, and the present
experimental upper limit \cite{2005PhRvL..95h1101F} is still far
from the expectations on the secondary antideuteron flux which are produced by 
spallation of cosmic rays on the interstellar medium 
\cite{1997PhLB..409..313C,2005PhRvD..71h3013D}, but {in fact} perspectives for the near
future are very encouraging.

In the present Paper we update {and improve our} calculation of the antideuteron primary
flux in a full two--zone diffusion model, consistent with a number of independent
cosmic ray  (CR) measurements, and {explicitly} estimate {the uncertainties
which affect the signal determination}. In addition, we also provide a new determination of the secondary
component. In Sect.~\ref{s:transport}, the framework and ingredients for the solution 
of the two-zone transport equation are recalled. In Sect.~\ref{sec:production},
the coalescence model for the nuclear fusion process 
is discussed both for secondary and primary \dbar\ production.
Sect.~\ref{secondary} is dedicated to the secondary \dbar\ flux and to the possible 
uncertainties affecting its evaluation. In Sect.~\ref{primary} the production of 
antideuterons from DM particles is detailed and results on the propagated fluxes 
are presented, together with the estimation of the uncertainties due to propagation, 
to the dark matter halo profile and to the DM annihilation final states. 
Finally, Sect.~\ref{discussion} demonstrates that antideuterons are probably one of the most
powerful dark matter indirect detection channel, and shows the optimistic potentials of 
next-to-come {balloon and} space based missions. We finally draw our conclusions in
Sect. \ref{sec:conclusions}.

\section{The propagation model}
\label{s:transport}

Cosmic ray fluxes are determined by the transport equation
as given, e.g., in \citet{1990acr..book.....B}. If steady-state
is assumed, the transport equation for any nuclear species can be rewritten in terms of the
cosmic ray differential density $dn(\vec{r})/dE\equiv N(\vec{r})$ as:
\begin{eqnarray}
  -\! \vec{\nabla} \!\left[ K\vec{\nabla} N(\vec{r}) \!-\! \vec{V_c} N(\vec{r})\right]
  \!-\! \frac{\partial}{\partial E}\!\left[
  \!- f_{\rm o} N(\vec{r}) \!+\! s_{\rm o}\frac{\partial N(\vec{r})}{\partial E} \right]\!\!\!\!\!\!\!\!\!\!\nonumber \\
= Q_{\rm source}(\vec{r}) - n(\vec{r})v\sigma_{\rm ine} N(\vec{r})\,.~~~~~~~~~~~~~~
  \label{eq:transport_CR}
\end{eqnarray}
The l.h.s describes the spatial diffusion ($K$) and convection ($V_c$), 
and the energy transport
(first and second order terms). The r.h.s corresponds to the primary, secondary
and tertiary (only for antinuclei) source terms, and the sink (spallative destruction) for
the  considered species. 
The most general form for the diffusion coefficient is $K(E,\vec{r})$. Energy
gain and losses depend on $(E,\vec{r})$ as well. The first order
term $f_{\rm o}(E,\vec{r})$ corresponds to the sum of four contributions:
ionization, Coulomb and adiabatic losses, and first order reacceleration.
Ionization losses take place in the neutral interstellar medium (ISM), while
the Coulomb ones in the completely ionized plasma, dominated by scattering off
the thermal electrons~\citep{1994A&A...286..983M,1998ApJ...493..694M}. 
The spatial dependence of these two terms is encoded
in the distribution of the neutral and ionized gas.
Adiabatic losses are due to the expanding wind and their spatial dependence is
related to the gradient of $\vec{V_c}$. The last contribution to $f_{\rm o}(E,\vec{r})$
has the same origin as the second-order term $s_{\rm o}(E,\vec{r})$. Those come from
the scattering of the charged particles off the turbulent magnetic fields of the Galaxy.
As well as being responsible for spatial diffusion, the Alfv\'enic waves also lead to
energy drift and reacceleration. A minimal reacceleration scheme is well-motivated
\citep{1994ApJ...431..705S} and allows to calculate the $f_o$ and $s_o$ coefficients. Similar
albeit more empirical forms have also been used \citep{1995ApJ...441..209H,2001ApJ...555..585M}.
In all these models, the strength of the reacceleration is mediated via the Alfv\'enic speed
$V_a$ of the scatterers.


  \subsection{The two-zone disk-halo model}
A full numerical treatment is generally required to solve the transport equation, as described,
e.g., in \citet{1998ApJ...509..212S}. However, analytical (or semi-analytical) solutions may be
derived assuming a simplified description of the spatial dependence of some
parameters in Eq.~(\ref{eq:transport_CR}). 
The two-zone diffusion model  \citep{1969ocr..book.....G,1980Ap&SS..68..295G}, 
based on the description of the Galaxy as a thin gaseous disk embedded in a thick diffusive halo,
proved to be successful in reproducing the nuclear  \citep{1992ApJ...390...96W,2001ApJ...555..585M,
2002A&A...381..539D}, antiproton  \citep{2001ApJ...563..172D} and radioactive isotopes data
\cite{2002A&A...381..539D}. It also allows to
treat contributions from dark matter (or other exotic) sources
located in the diffusive halo \citep{2004PhRvD..69f3501D,2005PhRvD..72f3507B,2005PhRvD..71f3512S},
which is the aim of this Paper. We remind below the salient features of this model,
which has been extensively detailed in Refs. \cite{2001ApJ...555..585M,2002A&A...394.1039M,revue}. 

    \paragraph{Geometry ($L$, $R$ and $h$).} The Galaxy is defined as a cylinder with a
diffusive halo of  half-height $z=L$  and radius $r=R$. The halo thickness $L$ is a free
parameter of the model. The interstellar (IS)  gas and the nuclei accelerators are contained in a
thin-disk of half-height $h\ll L$. The two parameters $h$ and $R$ are set to
100~pc and 20~kpc, respectively.
 
    \paragraph{Diffusion coefficient ($K_0$ and $\delta$).}
Diffusion arises because charged particles interact with the galactic magnetic field
inhomogeneities. The diffusion coefficient $K(\vec{r},E)$ 
is related to the power spectrum of these
inhomogeneities, which is poorly known. Several analytical forms for $K$ have been assumed in
the literature. We consider here the standard rigidity (${\cal R}=pc/Ze$) dependent form 
$K(E) = \beta \, K_0 \times {\cal R}^\delta$,
where the normalization $K_0$ is expressed in units of kpc$^2$~Myr$^{-1}$.
The same diffusion coefficient is assumed throughout the Galaxy,
i.e. in the disk and in the halo. $K_0$ and $\delta$ are free parameters of the model. 

    \paragraph{Galactic wind and Alfv\'enic speed ($V_c$ and $V_a$).} 
The convective wind is assumed to be of constant magnitude directed 
outwards perpendicular to the Galactic plane $\vec{V_c}=V_c \vec{e}_z$ .
The reacceleration strength, mediated by the Alfv\'en velocity
$V_a$,  is confined to the thin disk.
The first and second order terms $f_{\rm o}$ and $s_{\rm o}$ in Eq.~(\ref{eq:transport_CR})
follow the formulation given in Ref. \citep{2001ApJ...555..585M}.

\begin{table}
\[
\begin{array}{|c|c|c|c|c|c|c||c|c|} \hline
{\rm case} &  \delta  & K_0 & L & V_c & V_A & \chi^2_{\rm B/C}  \\
  & & {\rm (kpc^2/Myr)} & ({\rm kpc}) & ({\rm km/s}) & ({\rm km/s}) & \\\hline \hline
{\rm max} &  0.46  & 0.0765 & 15 & 5   & 117.6 & 39.98 \\
{\rm med} &  0.70  & 0.0112 & 4 & 12   &  52.9 & 25.68 \\
{\rm min} &  0.85  & 0.0016 & 1 & 13.5 &  22.4 & 39.02 \\
\hline \hline
\end{array} 
\]
\caption{Transport parameters {providing} the maximal, median and
minimal primary antideuteron flux and compatible with B/C analysis
($\chi^2_{\rm B/C}<40$) \cite{2001ApJ...555..585M,2002A&A...394.1039M}.}
\label{table:prop}
\end{table}
The free  parameters of this propagation model|the diffusive halo size $L$, 
the normalization $K_0$ and the slope $\delta$ of the diffusion coefficient,
 the value of the constant galactic wind $V_c$ and
the level of reacceleration through the Alfv\'enic speed $V_a$|were
constrained from the study of the B/C ratio
in Refs. \cite{2001ApJ...555..585M,2002A&A...394.1039M}.
In this Paper we take advantage of the parameters found in Ref.~\cite{2001ApJ...555..585M}
and listed in Table \ref{table:prop}.
Actually, when fitting to existing B/C data, a strong degeneracy of the
transport parameters is observed, meaning that many sets of these five parameters
are acceptable and lead to the same B/C ratio, but also to the
same secondary (standard) antiproton flux \citep{2001ApJ...563..172D}.
This degeneracy is broken for sources located in the diffusive
halo, leading to large astrophysical uncertainties for the relevant 
fluxes \citep{2002A&A...388..676B,2003A&A...398..403B,2004PhRvD..69f3501D}.
The same conclusions are {obtained here} for antideuterons. We will come back to this
point in Secs.~\ref{sec:secondary} and \ref{sec:primary}.
%

\subsection{The case of antideuterons\label{sec:sources}}
Once the astrophysical framework for the transport of (anti)nuclei is set,  the calculation
of the antideuteron flux rests on the \dbar\ specificities regarding the source term and its
nuclear interactions [r.h.s. of Eq.~(\ref{eq:transport_CR})].
The source term for antinuclei is
usually cast in separate contributions:
\begin{equation}
Q_{\rm source}(\vec{r},E)= Q_{\rm prim}(\vec{r},E) + Q_{\rm sec}(\vec{r},E) + 
Q_{\rm ter}(\vec{r},E)\,.
\end{equation}
Among these three terms, only primary and secondary
are {\em true} sources. 
They will be discussed in Secs.~\ref{sec:secondary}
and~\ref{sec:primary}.

The tertiary term was emphasized in Ref.~\cite{1999ApJ...526..215B} to
describe the process corresponding to the non-annihilating
interaction of a CR antinucleus with an atom of the IS gas.
In a medium of constant density $n$ ($v$ is the CR velocity):
\begin{eqnarray}
Q_{\rm ter}(\vec{r},E) & = &
{\displaystyle \int_{E}^{+ \infty}}
\!\!\!\! n v' \,{\displaystyle
\frac{d \sigma^{\rm non-ann}_{\bar{d} H \to \bar{d} X}}{dE}}
\left\{ E' \! \to \! E \right\} \, N(\vec{r}, E')\; dE'
\nonumber \\
& - & \;\;\;\;\; n v \,\; \sigma^{\rm non-ann}_{\bar{d}H \!\to \! \bar{d}X}
\left\{ E \right\} \;N(\vec{r}, E)\;.
\label{eq:tertiary}
\end{eqnarray}
The cross sections in Eq.~(\ref{eq:tertiary}) refer to inelastic non-annihilating
processes and are detailed in the Appendix. 
The tertiary mechanism does not actually create new antideuterons. It merely
states that the number of antinuclei observed at energy $E$ has to take
into account the  redistribution of those with energy $E'>E$ (first term, positive contribution),
minus the total number of \dbar\ redistributed to lower energies
(second term, negative contribution). The tertiary contribution is treated as a corrective
factor and dealt with iteratively: the equilibrium spectrum $N^{(0)}(\vec{r},E)$
is first calculated with $Q^{(0)}_{\rm ter}\equiv 0$, then $Q^{(1)}_{\rm ter}$ calculated
with $N^{(0)}(\vec{r},E)$ in Eq.~(\ref{eq:tertiary}) to obtain $N^{(1)}(\vec{r},E)$, etc.
For antideuterons, due to the small cross section|as shown in
Ref.~\citep{2005PhRvD..71h3013D}|only one iteration is necessary to converge to
the solution (compared to a few iterations for antiprotons \citep{2001ApJ...563..172D}).

Hence, whether secondary or primary (or a mixture of both)
sources are considered, three cross sections always enter the calculation:
the differential non-annihilating inelastic cross section $d\sigma^{\rm non-ann}/dE$,
the total non-annihilating inelastic cross section $\sigma^{\rm non-ann}$
and the total annihilating inelastic cross section $\sigma^{\rm ann}$
[appearing in the r.h.s. of Eq.~(\ref{eq:transport_CR})]. Considering the ISM
as a mixture of H and He, the full calculation requires six cross sections.
 Our calculations, based on
the parameterizations discussed at length in \citet{2005PhRvD..71h3013D}, are discussed
in the Appendix where, in particular, we update $d\sigma^{\rm non-ann}/dE$.
Note however, that even if some of these cross sections are slightly modified, the
impact on
the propagated spectra do not change the conclusions found in Ref. \citep{2005PhRvD..71h3013D}.

In the two-zone diffusion/convection/reacceleration model,
it is possible to extract semi-analytical solutions of 
Eq.~(\ref{eq:transport_CR}), based on Bessel
expansions of the transport equation. We do
not wish to repeat the various steps of this derivation, nor to rewrite
the complete form of the solutions, which have already been given in
several papers. The solution for the so-called antideuteron standard source (of secondary
origin in the galactic disk) and the numerical procedure to treat reacceleration
is detailed in Ref.~\citep{2001ApJ...563..172D}. The solution for an exotic source distributed
in the whole diffusive halo of the Galaxy 
can be found in Ref.~\cite{2002A&A...388..676B,2004PhRvD..69f3501D}.
Actually these two papers refer to \pbar, but formally, the solutions apply
to \dbar\ as well.

\section{Antideuteron production}
\label{sec:production}

The production of cosmic antideuterons is based on the fusion process of a \pbar\
and \nbar\ pair. One of the simplest but powerful treatment of the fusion of two or
more nucleons is based on the so--called coalescence model which, despite its
simplicity, is able to reproduce remarkably well the {available} data on light nuclei and
antinuclei production in different kinds of collisions. 
In the coalescence model, the momentum distribution of the (anti)deuteron
is proportional to the product of the (anti)proton and (anti)neutron momentum
distribution \citep{1963PhRv..129..836B,1963PhRv..129..854S}.
That function depends on the difference $\Delta_{\vec{k}}$
between (anti)nucleon momenta. It is strongly peaked around
$\Delta_{\vec{k}}\simeq \vec{0}$ (compare the minimum energy to form
a \dbar, i.e. $4m_p$, with the binding energy $\sim 2.2$~MeV), so that
\begin{equation}
\vec{k}_{\bar{p}}\simeq \vec{k}_{\bar{n}} \simeq \frac{\vec{k}_{\bar{d}}}{2}\;.
\end{equation}
The \dbar\ density in momentum space is thus written as
the \pbar\ density times the probability to find an $\bar{n}$
within a sphere of radius $p_0$ around $\vec{k}_{\bar{p}}$
(see, e.g. Ref.~\citep{1994ZPhysC..61...683}):
\begin{equation}
\gamma \frac{d{\cal N}_{\bar{d}}}{d\vec{k}_{\bar{d}}} = \frac{4\pi}{3} p_0^3
  \cdot    \gamma \frac{d{\cal N}_{\rm \bar{p}}}{d\vec{k}_{\bar{p}}}
  \cdot    \gamma \frac{d{\cal N}_{\rm \bar{n}}}{d\vec{k}_{\bar{n}}}\;.
  \label{eq:coal}
\end{equation}
The coalescence momentum $p_0$ is a free parameter
constrained by data on hadronic production.
Note that the coalescence model has been refined to account for heavy nuclei
reactions \citep[see, e.g.][]{1979PL..B85...38,1988JPhG...14..937D}, but
as it is not relevant for this study, we will stick to the simple Eq.~(\ref{eq:coal}).

The number $d{\cal N}^{\rm R}_{\rm X}$ of particles ${\rm X}$
produced in a single reaction $R$ and which momenta are $\kX$
can be expressed as a function of the total available energy
$\sqrt{s}$, the inclusive (i.e. total inelastic or reaction cross section)
and the differential  cross section:
\begin{equation}
d{\cal N}^{R}_{\rm X}  =  {\displaystyle \frac{1}{\sigma_{\rm inel}^{\rm R}}} \,
{d^{3} \sigma_{\rm X}}(\sqrt{s} , \kX) \;.
\label{eq:Npar}
\end{equation}
For instance, in our specific case $\rm X$ are the antinucleons and
antideuterons created in the  $pp$, $p$He and HeHe reactions between
$p$-He CRs and  H-He in the ISM.
Assuming the usual equality between the unmeasured \nbar\ and the 
measured \pbar\ cross sections, and combining the two previous expressions
Eqs.~(\ref{eq:coal}) and (\ref{eq:Npar}) we get:
\begin{equation}
E_{\bar{d}} \frac{d^{3} \sigma^{\rm R}_{\bar{d}}}{d\vec{k}_{\bar{d}}}
 =
    \frac{1}{\sigma_{\rm inel}^{\rm R}}
  \cdot  
    \frac{4\pi}{3} p_0^3
  \cdot  
    \frac{m_{\bar{d}}}{m_{\bar{p}}^2}
  \cdot    
    \left( E_{\bar{p}}\frac{d\sigma^{\rm R}_{\rm \bar{p}}}{d\vec{k}_{\bar{p}}} \right)^2\;.
\label{eq:BIGcoal}
\end{equation}

The hypothesis of factorization of the probabilities is fairly well
established from experiments at high energies \citep[see,
e.g.,][]{2005PhRvD..71h3013D}.  For spallation reactions, however, the
bulk of the antiproton production takes place for an energy $\sqrt{s}
\sim 10$ GeV, which turns out to be of the same order of magnitude as
the antideuteron mass. Pure factorization should break in that case as
a result of energy conservation. Two ansatz have been used in order to
correct that effect for this regime: in Ref.~\citep{1997PhLB..409..313C,2000PhRvD..62d3003D} it was assumed that,
while the  first antinucleon is produced with $\sqrt{s}$,  the center
of mass energy available for the production of the second antinucleon
is reduced by twice the energy carried away by the first
antinucleon. Instead, in \citet{2005PhRvD..71h3013D} the threshold
production was phenomenologically taken into account through an $A+2$
phase space factor. The latter description seems more appropriate as,
while preserving the correct asymptotic properties, it does not favor
any mechanism for the pair production \citep{2004TheseRemy}.

The coalescence momentum $p_0$ is linked to the measured coalescence factor
$B_{A=2}$ (hereafter simply $B_2$):
\begin{eqnarray}
B_2 &\equiv& 
  \sigma_{\rm inel}^{\rm R} \cdot E_{\bar{d}} \frac{d^{3}
    \sigma^{\rm R}_{\bar{d}}}{d\vec{k}_{\bar{d}}} \cdot
   \left( E_{\bar{p}}\frac{d\sigma^{\rm R}_{\rm \bar{p}}}{d\vec{k}_{\bar{p}}} \right)^{-2}\;,
\end{eqnarray}
so that
\begin{equation}
p_0 = \left( \frac{1}{B_2}\cdot \frac{m_{\bar{d}}}{m_{\bar{p}}^2}\cdot \frac{4\pi}{3}\right)^{-1/3}\; .
\label{eq:B2_to_p0}
\end{equation}
The $B_2$ coefficient has been measured for proton-proton, proton-nucleus
and heavy ion collisions (see a summary and references in Refs.
\citep{2004EJPh..C36...413,2005PhRvD..71h3013D}).
More recently, several other channels have also been measured at high energy:
photo-production \citep{2004EJPh..C36...413}, DIS production \citep{2007NPhB..786..181}
and $e^+e^-$ production at the $Z$ \citep{2006PhLB..639..192A} and $\Upsilon(1S)$
\citep{2007PhRvD..75a2009A} resonances. The $e^+ e^-$ channel is of particular
interest for the DM annihilation reactions.

\subsection{Hadronic production}

For the hadronic processes, the coalescence momentum can
directly be fitted to data. However,  different
assumptions regarding the set of data to retain can
lead to different values of $p_0$. Note that many recent data are
available for $A+A$ systems, in addition to $pp$ and $pA$ reactions.
However, the mechanisms at play in heavy ion collisions
are not necessarily those of lighter systems
(see, e.g., discussion in Sect.~II.A of Ref.~\citep{2005PhRvD..71h3013D} and the
results of Ref.~\citep{2003NJPh....5..150A}),
 so that these reactions are discarded in the rest of our analysis.

\citet{1997PhLB..409..313C} used $pp$ data from Refs.
\citep{1975PNu..B97..189,1978NCL..21...189,1987SJNP..45...845}
and $pA$ collisions from Refs. \citep{1987SJNP..45...845} 
(see Fig.~1 in Ref.~\citep{1997PhLB..409..313C}) to derive
a coalescence momentum $p_0=58$~MeV. Based on kinematical relevance of
the measured reactions, these authors disfavored the $pp$ data from Ref.~\citep{1978NCL..21...189}, but underlined that a value $p_0=75$~MeV, compatible
with their whole set of data, would merely provide
twice as more antideuterons.

In \citet{2005PhRvD..71h3013D}, a larger set of data is used,
including many $pA$ reactions (see their Tab.~I and
references therein). The approach is more sound since a $\chi^2$
analysis on the momentum distribution of the fragments was performed,
taking also into account the phase space. This leads
to an estimate of $p_0=79$~MeV.
Note that at variance with the choice of \citet{1997PhLB..409..313C},
Duperray et al. discarded the data from Ref.~\citep{1987SJNP..45...845} because
they give a poor $\chi^2$ value compared to all other data. It is thus not 
surprising that these authors end up with a value close
to $p_0=75$~MeV quoted in \citet{1997PhLB..409..313C}.
In the present Paper, we take directly the cross sections derived in \citet{2005PhRvD..71h3013D}
using the value $p_0=79$~MeV. 

\subsection{Weak production}

At LEP energies, (anti)deuteron production occurs through $e^+e^-$ annihilations into
$q\bar{q}$ pairs, a mechanism similar to the \dbar\ production in DM annihilation reactions.
Based on theoretical arguments, it has been argued \citep{1994ZPhysC..61...683}
that the antideuteron  yields in $e^+e^-$ reactions should be smaller than in hadronic reactions.
However, the ALEPH Collaboration \citep{2006PhLB..639..192A} has found that this theoretical
prediction (see Fig.~5 in ALEPH paper) underestimates their measured \dbar\ inclusive cross
section. They derive (see their Fig.~6)  a value $B_2=3.3\pm 0.4\pm0.1 \times 10^{-3}$~GeV$^2$
at the $Z$ resonance, which translates into $p_0=71.8\pm 3.6$~MeV, very close to
the $p_0=79$~MeV derived for the hadronic production. Hence,
in the remaining of the Paper, the value of $p_0=79$ MeV will be retained for
both the processes of hadronic and electroweak origin. 


\section{Secondary antideuterons\label{sec:secondary}}
\label{secondary}

Secondary antideuterons are produced in the galactic disk from the collisions of 
cosmic protons and helium nuclei over the ISM.  We evaluate here the \dbar\
propagated fluxes as well as the nuclear and propagation uncertainties, similarly
to what was done for \pbar\ in Ref.~\citep{2001ApJ...563..172D}. 

\subsection{Median flux}

\begin{figure}[t]
\centering
\vspace{-42pt}
\includegraphics[width=1.1\columnwidth]{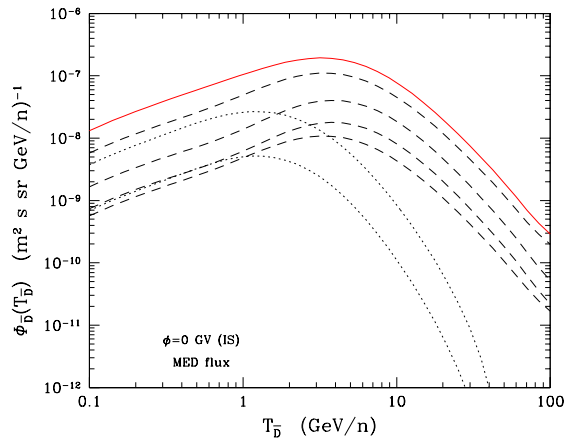}
\vspace{-30pt}
\caption{Contribution of all nuclear channels to the \dbar\  secondary flux. Dashed
lines, from top to bottom refer to: p+H, p+He, He+H, He+He. Dotted lines, from top to
bottom stand for: \pbar+H, \pbar+He. Solid line: sum of all the components.}
\label{fig:separated}
\end{figure}

The secondary \dbar\
flux is the sum of the six contributions corresponding to $p$, He and \pbar\ cosmic
ray fluxes impinging on H and He  IS gas (other reactions are negligible
\citep{2005PhRvD..71h3013D}). The p and He fluxes were fitted on BESS 
\citep{2000ApJ...545.1135S} and AMS 
\citep{2000PhLB..472..215A,2000PhLB..490...27A,2000PhLB..494..193A} high energy
data with a power law {spectrum} (see details in Ref.~\citep{2001ApJ...563..172D})
$\Phi (\rm T) = N\,(\rm T/GeV/n)^{-\gamma}$.
The best fit corresponds to $N_{\rm p} = 
13249$~m$^{-2}$~s$^{-1}$~sr$^{-1}$~(GeV/n)$^{-1}$
and $\gamma_{\rm p} = 2.72$, and $N_{\rm He} = 
721$~m$^{-2}$~s$^{-1}$~sr$^{-1}$~(GeV/n)$^{-1}$
and $\gamma_{\rm He} = 2.74$.
The uncertainty on these two fluxes is small and leads to
negligible uncertainties in the \pbar\ and \dbar\ spectrum.
Contributions to the \dbar\ flux from $\bar{p}+H$ and $\bar{p}+{\rm He}$
reactions are evaluated using the \pbar\ flux calculated in the same
run. The production cross sections for these specific processes
are those given in Ref.~\citep{2005PhRvD..71h3013D}.

The different contributions to the total secondary antideuteron flux, 
calculated for the best fit propagation configuration 
(the ``med" one in Table \ref{table:prop}), i.e. $K_0 =
0.0112$~kpc$^2$~Myr$^{-1}$, $L = 4$~kpc, $V_c = 10.5$~km~s$^{-1}$  and $V_a =
52.1~$km~s$^{-1}$, are shown in Fig.~\ref{fig:separated}. 
As expected, the dominant production channel is the one from $p$-$p$ collisions,
followed by the one from cosmic protons on IS helium (p-He).  As shown in Ref.~\citep{2005PhRvD..71h3013D}, the \pbar+H channel is dominant at low energies,
and negligible beyond a  few GeV/n. The effect of
energy losses, reacceleration and  tertiaries add up to replenish the low
energy {tail}. The maximum of the flux reaches the value of  $2\cdot 10^{-7}$  particles 
(m$^2$ s sr GeV/n)$^{-1}$ at 3-4 GeV/n. At 100 MeV/n it is decreased by 
an order of magnitude, thus preserving an  interesting window for possible
exotic contributions characterized by a flatter spectrum.

\subsection{Propagation uncertainties}
\begin{figure}[t]
\centering
\vspace{-42pt}
\includegraphics[width=1.1\columnwidth]{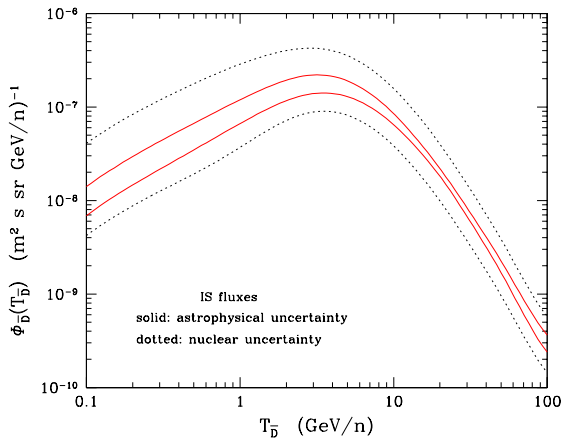}
\vspace{-30pt}
\caption{Dominant uncertainties on the interstellar secondary \dbar\ flux. 
Solid lines: propagation uncertainty band. 
Dotted lines: nuclear uncertainty band.}
\label{sec_uncert}
\end{figure}

For {a determination} of the propagation uncertainties, we follow the same approach
as in Ref.~\citep{2001ApJ...563..172D}.  We
calculate the secondary antideuteron flux for all the propagation parameter 
combinations providing an acceptable fit to stable nuclei 
\citep{2001ApJ...555..585M}. 
 The resulting envelope for the secondary antideuteron flux is presented in Fig.~\ref{sec_uncert}.
 The solid lines  delimit the uncertainty band due to the
degeneracy of the propagation parameters: at energies below {1--2 GeV/n}, the uncertainty 
is 40-50 \% {around }the average {flux}, while at 10 GeV/n it decreases to 
 $\sim 15$ \%. This behavior is {analogous to that obtained} for \pbar\ \citep{2001ApJ...563..172D}
and is easily understood. The degenerate transport parameters combine to give
the same grammage in order to reproduce the B/C ratio. Indeed, the 
grammage crossed by C to produce the secondary species B is also
crossed by $p$ and He to produce the secondary \pbar\ and \dbar. In short,
a similar propagation history associated with a well constrained B/C
ratio explains the small uncertainty. With better measurement of B/C
expected soon, e.g. from PAMELA \citep{2007APh....27..296P} or TRACER \citep{Ave:2008as},
this uncertainty will further decrease and could become negligible.

\subsection{Nuclear uncertainty}
The possible nuclear uncertainty can arise from two different
sources. The first one is directly related to the elementary production process
$d\sigma^{\rm R}_{\rm \bar{p}}$.  It was found in Ref.~\citep{2001ApJ...563..172D} that
this may be cast  into a $\pm25\%$ in the \pbar\ propagated flux, so that it should
be translated to a rough $\pm50\%$ in the \dbar\ flux. Second, there is the
uncertainty on the coalescence momentum $p_0$.  Using an independent model (i.e. different
from the coalescence scheme) for
\dbar\ production, Ref.~\citep{2005PhRvD..71h3013D} found that, conservatively, the
\dbar\ background was certainly no more than twice the flux calculated with
$p_0=79$~MeV.

To some extent, these two uncertainties are correlated as the value
of $p_0$ depends on the choice for $d\sigma^{\rm R}_{\rm \bar{p}}$.  Hence, to be
very conservative and to keep a simple approach, we have spanned 
all hadronic production cross sections  in the range $^{+100\%}_{-50\%}$ around 
their reference value. If we wished to translate this into an uncertainty
on the coalescence momentum, this would lead to the {\em effective} range $p_0=79 ^{+26}_{-13}$~MeV.
Finally, in order to estimate the maximal flux with the most conservative attitude
(the highest is the secondary flux the lowest is the chance to outline an exotic
contribution), the non-annihilating cross section 
was doubled, as its value is probably only a lower limit (see Appendix). On the
other hand, to evaluate the minimal flux, we switched off the $\bar{p}+H(He)
\rightarrow \bar{d}+X$ contributions, which intensity remains very uncertain.

The dotted lines in Fig.~\ref{sec_uncert} take into account the sum of all the possible
uncertainties of nuclear source, as described above. At the lowest energies the
flux is uncertain by almost one order of magnitude, at 100 GeV/n by a factor of 4. 
We have checked that the  solar
wind mildly decreases the IS flux at low energies but leaves the uncertainty
magnitude unchanged. 
It is obvious from Fig.~\ref{sec_uncert} that the uncertainties on nuclear and hadronic
cross sections (dashed lines) are more important than the ones coming from the propagation
models (solid lines). We emphasize once more our conservative attitude in estimating
the nuclear band. However, if no dedicated campaigns of measurements for these
cross sections {will be} carried {out} in the future, these uncertainties are not likely to
be significantly reduced.

\section{Primary antideuterons\label{sec:primary}}
\label{primary}

The source term for primary \dbar\ to be cast into Eq.~(\ref{eq:transport_CR}) is:
\beq
q_{\bar{d}}^{\rm prim}(r,z,E)=\eta\, \xi^2 \sigmav_0 \,  \frac{dN_{\bar{d}}}{d E_{\bar{d}}} \,
\left(\frac{\rho_{\rm DM}(r,z)}{m_\chi}\right)^2\;,
\label{source_prim}
\eeq
where $\sigmav_0$ is the thermal average of annihilation cross section times the WIMP velocity, 
$dN_{\bar{d}}/d E_{\bar{d}}$ is the source spectrum,
$\rho_{\rm DM}(r,z)$ is the distribution of the DM in the Galaxy and $m_\chi$ is the WIMP mass. 
The quantity $\sigmav_0$ depends on the particle physics model. If not differently stated, we fix 
its value to $2.3 \cdot 10^{-26}$ cm$^3$ s$^{-1}$, which corresponds to a thermal CDM relic 
{able to explain the observed amount of cosmological dark matter 
\cite{Hinshaw:2008kr,Komatsu:2008hk,Dunkley:2008ie}. This will be our
reference value for most of the analysis.
The coefficient $\eta$ depends on the
particle being or not self--conjugate: for instance, for a fermion
it is $1/2$ or $1/4$ depending on whether the WIMP is a
Majorana or a Dirac particle. In the following we will adopt $\eta = 1/2$. The quantity
$\xi$ parameterizes the fact that the dark halo may not be
totally made of the species under scrutiny (e.g a neutralino or a sneutrino) when this candidate
possesses a relic abundance which does not allow it to be the dominant DM component (see e.g \cite{Bottino:2008sv}
or \cite{Bottino:1992wj}). In this case $\rho_{\chi} = \xi \rho_{\rm DM}$ with
$\xi < 1$. For our reference value for
$\sigmav_0$ clearly one has $\xi$=1. The DM candidate
may then be identified with a neutralino \citep{2004PhRvD..69f3501D} 
or a sneutrino \cite{Arina:2007tm} in various supersymmetric schemes, but for the purposes of our
discussion it does not need to be specified. We in fact wish to maintain the discussion at the
most general level: we just need to specify the final state particles produced in the DM
annihilation process and the ensuing energy spectra. The final--state particles all belong to
the Standard Model, and this allows us to perform our discussion on a totally general basis.
We will at the end specify our candidate to be the neutralino and discuss experimental capabilities
in the framework of some specific supersymmetric scheme.}

Below, we briefly recall the main steps for the calculation of the source term,
before focusing on the propagation of these antideuterons in the Galaxy (Sect.~\ref{sect:res}),
which is one of the main novelty in this Paper.

\subsection{Antideuteron source spectrum}

The production of antideuterons from the pair-annihilation of dark matter
particles in the halo of our Galaxy was proposed in \citep{2000PhRvD..62d3003D}.
The interest in this possible DM detection channel has been the physics case for the 
proposal of the GAPS experiment \citep{2002ApJ...566..604M,2006JCAP...01..007H,
2004NIMPB.214..122H} and it has also been considered in Refs.
\citep{2005JCAP...12..008B,Bottino:2007qg,dark2004}.

As previously discussed (see Sect.~\ref{sec:production}) the production 
of a \dbar\ relies on the availability of a \pbar\ -- \nbar\ pair in a single 
DM annihilation. 
In the case of a WIMP pair annihilation, the differential multiplicity
for antiproton production may be expressed as
\begin{equation}
{\displaystyle \frac{dN_{\bar{p}}}{d E_{\bar{p}}}} \; = \;
{\displaystyle \sum_{\rm F , h}} \, B_{\rm \chi h}^{\rm (F)} \,
{\displaystyle \frac{dN_{\bar{p}}^{\rm h}}{{d }E_{\bar{p}}}} \; .
\end{equation}
The annihilation into a quark or a gluon $h$ is realized through 
the various final states F with branching ratios $B_{\rm \chi h}^{\rm (F)}$.
Quarks or gluons may in fact be directly produced when a WIMP pair annihilates or 
they may alternatively result from the intermediate production of Higgs bosons or 
gauge bosons. Each quark or gluon $h$ then generates 
jets whose subsequent fragmentation and hadronization yield an antiproton
energy spectrum ${dN_{\bar{p}}^{\rm h}} / {d E_{\bar{p}}}$.

As in Ref.~\citep{2000PhRvD..62d3003D},
we assume that the probability to form an antiproton (or an antineutron)
with momentum $\vec{k}_{\bar{p}}$ ($\vec{k}_{\bar{n}}$), is essentially isotropic:
\beq
{\displaystyle \frac{dN_{\bar{p}}}{d E_{\bar{p}}}}(\chi + \chi \to \bar{p} + \ldots)
\; = \;
4 \pi \, k_{\bar{p}} \, E_{\bar{p}} \, {\cal F}_{\bar{p}}(\sqrt{s} = 2 m , E_{\bar{p}})
\; .
\eeq
Applying the factorization--coalescence scheme discussed above leads
to the antideuteron differential multiplicity
\beq
{\displaystyle \frac{dN_{\bar{d}}}{d E_{\bar{d}}}} = 
\left( {\displaystyle \frac{4 \, p_0^{3}}{3 \, k_{\bar{d}}}} \right)
\cdot
\left( {\displaystyle \frac{m_{\bar{d}}}{m^2_{\bar{p}}}} \right)
\cdot
{\displaystyle \sum_{\rm F , h}}  B_{\rm \chi h}^{\rm (F)} 
\left\{
{\displaystyle \frac{dN_{\bar{p}}^{\rm h}}{d E_{\bar{p}}}}
\left( E_{\bar{p}} = \frac{E_{\bar{d}}}{2} \right)
\right\}^{2} \; .
\label{dNdbar_on_dEdbar_susy}
\eeq
We assume, as discussed in Sect.~\ref{sec:production}, that the same value 
of the coalescence momentum $p_0=79$ MeV holds as for hadronic reactions.

The evaluation of the differential antiproton spectrum $dN_{\bar{p}}^{\rm h}/d E_{\bar{p}}$ 
follows the treatment of Ref.~\citep{2004PhRvD..69f3501D}. 
We refer to this paper for the details of the \pbar\ spectra from all the annihilation channels.
The resulting \dbar\ source spectra from different final states are not directly shown here. Instead,
we will provide examples of propagated \dbar\ spectra for the various final states in
Fig.~\ref{ann_state}.

\subsection{Dark matter halo profile}
\label{sec:DM}
The distribution of DM inside galaxies is a very debated issue 
(see e.g. Ref.~\citep{2008A&A...479..427L}
for a brief highlight on recent results and relevant references). 
Different analyses of rotational curves observed for several types 
of galaxies strongly favour a cored dark matter distribution, flattened towards 
the central regions (Ref.~\cite{2004MNRAS.353L..17D} and references therein). 
On the other side,
many collisionless cosmological N-body simulations in $\Lambda$-CDM models
are now in good agreement among themselves \citep{2007arXiv0706.1270H}, but for the
very central regions some resolution issues remain open. 
It has been recently stressed that asymptotic slopes may not be reached at
all at small scales \citep{2004MNRAS.349.1039N,2006MNRAS.365..147S,
2006AJ....132.2685M,2006AJ....132.2701G,2007ApJ...663L..53R}.
However, it is not clear whether the central cusp is steepened or flattened
when the baryonic distribution is taken into account 
(e.g. \citep{2006MNRAS.366.1529M,2006Natur.442..539M}).
For definiteness, we  consider a generic dark matter distribution:
\beq
\rho_\chi \equiv \rho_{\rm CDM}(r) = \rho_\odot \,
\left\{
{\displaystyle \frac{r_{\odot}}{r}} \right\}^{\gamma} \,
\left\{
{\displaystyle \frac{1 \, + \, \left( r_{\odot} / a \right)^{\alpha}}
{1 \, + \, \left( r / a \right)^{\alpha}}}
\right\}^{\left( \beta - \gamma \right) / \alpha} \; ,
\label{eq:profile}
\eeq
where $r_{\odot} = 8$ kpc is the distance of the Solar System from the galactic
center. The spherical pseudo-isothermal and 
cored DM profile with ($\alpha, \beta, \gamma$)=(2,2,0) will 
be the reference in our calculations. The total local|Solar System|CDM density
has been set equal to $\rho_\odot= 0.42$ GeV cm$^{-3}$, the core radius to $a$=4 kpc.
 This value
and the total local density may be varied in large intervals by maintaining
good agreement with observations. The antideuteron flux is very
sensitive to the local distribution of dark matter $\rho_\odot$, since it appears squared
in the determination of the flux, while it is less sensitive to the chosen dark matter
distribution function (as was already underlined in Ref.~\citep{2004PhRvD..69f3501D} for \pbar).
For completeness and for comparison, we also consider some of the profiles obtained from 
$\Lambda$-CDM simulations:
i) a standard NFW profile having ($\alpha,\beta,\gamma$)=(1,3,1),
 with $a=21.746$~kpc \citep{1997ApJ...490..493N},
ii) the steeper DMS-1.2 (1,3,1.2) profile with $a=32.62$~kpc \citep{2004MNRAS.353..624D},
iii) and the modified NFW profile with an logarithmic slope (hereafter N04), with 
$a=26.4$~kpc  \cite{2004MNRAS.349.1039N} 
(similar to the Einasto profile \citep{2006AJ....132.2685M}).
Scale radii for NFW and DMS-1.2 profiles are taken from 
Ref.~\citep{2004PhRvD..70j3529F}, while the parameters for the N04 DM density 
distribution are the same as in Ref.~\citep{Bottino:2004qi}.
All these profiles are normalized to $\rho_\odot= 0.42$ GeV cm$^{-3}$, in order to isolate
the effect of the local density, which can be easily rescaled in the flux evaluation. 
We do not include any boost factor due to halo substructures. 
This conservative attitude is corroborated by the results of  \citet{2008A&A...479..427L}, 
where it has been shown that the boost factor is typically close to unity: only
for some extreme and unlikely configuration it can reach a factor of 10.

\subsection{Primary antideuteron flux and uncertainties}
\label{sect:res}

\begin{figure}[t] 
\centering 
\vspace{-42pt}
\includegraphics[width=1.1\columnwidth]{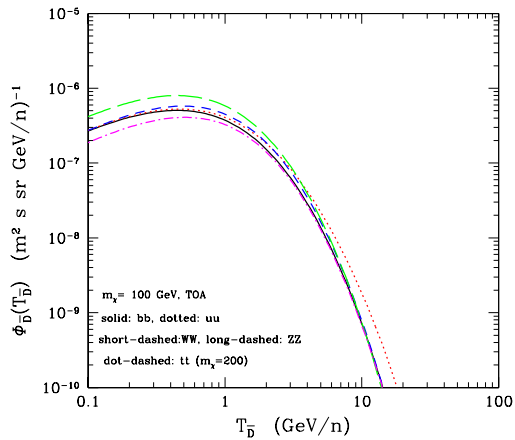} 
\vspace{-30pt}
\caption{Antideuteron flux for a WIMP mass $m_\chi$=100 GeV annihilating
into different final states: solid (black) line refers to $\bar{b}b$, 
 dotted (red) to $\bar{u}u$, short-dashed (blue) to $WW$, long-dashed 
 (green) line to $ZZ$. The dot-dashed (magenta) refers to $\bar{t}t$ and $m_\chi$=200 GeV.
 The annihilation cross section (here and in the following figures) is fixed at the value: 
 $\sigmav_{0} = 2.3 \cdot 10^{-26}$ cm$^3$ s$^{-1}$.} 
\label{ann_state} 
\end{figure}

\begin{figure}[t] 
\centering 
\vspace{-42pt}
\includegraphics[width=1.1\columnwidth]{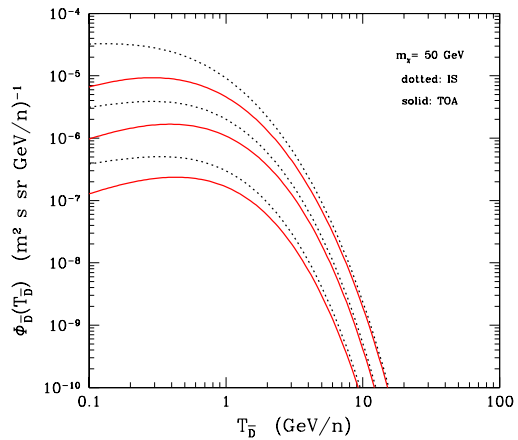} 
\vspace{-30pt}
\caption{Antideuteron flux for WIMPs of $m_\chi$=50 GeV. 
Dotted (black) lines refer to the interstellar flux, solid (red) lines stand for the
top--of--atmosphere flux, 
modulated at solar minimum. For each set of curves, the three lines refer to
the maximal, median and minimal propagation configurations defined in Table
\ref{table:prop}.} 
\label{prop} 
\end{figure}

\begin{figure}[t] 
\centering 
\vspace{-42pt}
\includegraphics[width=1.1\columnwidth]{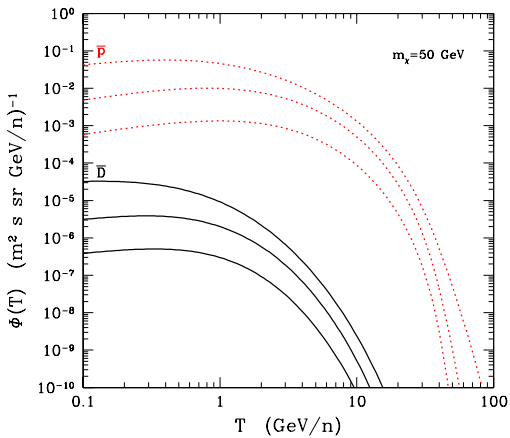} 
\vspace{-30pt} 
\caption{Uncertainty due to propagation models on the antideuteron (black solid
lines) and antiproton (red dotted lines) interstellar fluxes. The WIMP mass has been
fixed at the value $m_\chi$=50 GeV. For each set of curves, the three lines refer
to the maximal, median and minimal propagation configurations defined in Table
\ref{table:prop}.}
\label{pbar_dbar} 
\end{figure} 

\begin{figure}[t] 
\centering 
\vspace{-42pt}
\includegraphics[width=1.1\columnwidth]{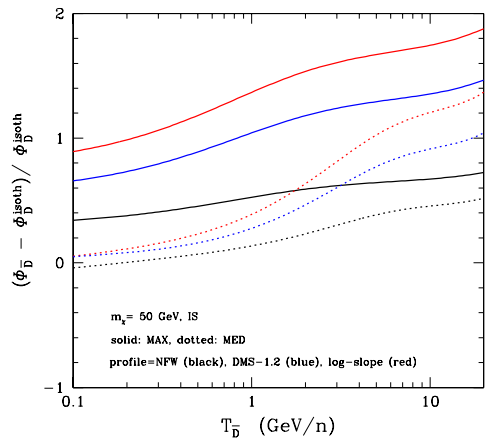} 
\vspace{-30pt} 
\caption{ Effect of changing the DM halo density profile, for a $m_\chi$=50 GeV WIMP
and for the ``max'' (solid) and ``med'' (dotted) configurations of Table \ref{table:prop}.
The effect is shown as the relative change in the IS antideuteron 
flux as compared with the reference case of a cored isothermal profile.
The lower (black) lines refer to the NFW profile \cite{1997ApJ...490..493N},  
the median (blue) lines to a cuspy profile with 1.2 slope 
\cite{2004MNRAS.353..624D} and the upper (red) ones to 
the N04 profile of Ref.~\cite{2004MNRAS.349.1039N}.}
\label{profiles} 
\end{figure}

In the present Section, we show our results for the propagated 
antideuteron flux from DM annihilation. We follow the prescriptions
detailed in the previous Sections for the production and the propagation
of antideuterons.
Top--of--atmosphere (TOA) fluxes are derived from the interstellar (IS) 
ones treating the effect of the solar modulation with the force
field approximation. If not differently stated, TOA fluxes correspond
to a solar minimum activity  with modulation potential $\phi$=0.5 GV.
For the reference propagation configurations, we {refer} the reader to 
Table \ref{table:prop}.

\subsubsection{Fluxes for various annihilation states}

Figure~\ref{ann_state} displays the \dbar\ flux from a WIMP of mass $m_\chi=100$ GeV.
Each curve corresponds to different pure ($i.e.$ with BR=1) annihilation final
states: $\bar{b}b$,  $\bar{u}u$, $WW$, $ZZ$ and $\bar{t}t$  (for which $m_\chi=200$ GeV). 
The aim of the figure is to show the effect on the observable flux of the different 
$\chi$-$\chi$ annihilation
final states from which the \dbar\ originate. 
The transport parameters are the ``med'' ones of Tab.~\ref{table:prop}
(providing the best fit to B/C data) and the fluxes are not modulated (IS spectra).
In the low energy part of the spectrum -- around and below 1 GeV/n -- 
it turns out that these fluxes show quite similar shapes {and comparable normalization} 
when varying the final state.
This  energetic range is the one in which a primary flux might emerge from 
the secondary counterpart. In addition, as we will also discuss at the end of our 
Paper, it is the window explorable by experiments in  a near future. 
For these reasons,  we will adopt the  antideuteron yield from an annihilation into
a pure $\bar{b}b$ final state as a simple but representative case for our 
discussions.

\subsubsection{Propagation uncertainties}

Fig.~\ref{prop} shows the uncertainties on the primary \dbar\ flux due
to propagation parameters. The three curves (dashed line: IS fluxes; solid lines: TOA fluxes)
correspond to the maximum, median and minimal set of propagation parameters as gathered in Table
\ref{table:prop}. 
The band between the upper and lower curve
estimates the uncertainty due to propagation. At the lowest energies
of  hundreds of MeV/n the total uncertainty reaches almost 2 orders of
magnitude,  while at energies above 1 GeV/n it is about a factor of
30.  The figure refers to a WIMP mass of 50 GeV but the results are
insensitive to this parameter, as well as from the solar modulation.
The magnitude of the propagation uncertainty is similar to the one
affecting the primary antiprotons \citep{2004PhRvD..69f3501D}, as 
explained in Fig.~\ref{pbar_dbar}.
 This behavior is drastically different from that observed  on the secondaries
(see Fig.~\ref{sec_uncert}). Indeed, their propagation history is very
different. Whereas secondaries originate from standard sources in the thin disk
of the Galaxy, exotic primaries are produced in all the diffusive halo of the
Galaxy.  As shown in Ref. \citep{2006astro.ph..9522M}, these primary antinuclei do
not suffer large energy losses, reacceleration or tertiary redistribution as
they rarely cross the thin disk. 
Most of them  arrive at Earth|substantially unshifted in energy|from an
effective  diffusion cylinder of height $L^*=\min(L,2K/V_c)$ and radius of a few
$L^*$ centered on the observer \citep{2003A&A...404..949M}.  Hence, the
parameters driving the uncertainty on the primaries are, at high energy, the
range allowed for the halo size $L$ (see Tab~\ref{table:prop}), whereas, at
lower energy, this uncertainty is further increased by the effect of the
galactic wind ($2K/V_c$ becomes smaller than $L$).

When comparing in details the \pbar\ and the \dbar\ fluxes from
Fig.~\ref{pbar_dbar}, the following conclusions can be drawn.
First, the antiproton fluxes are a factor of $10^4$ higher than the
antideuteron ones, as expected from the fusion process into \dbar.  
Then, at high energies the difference between the two
fluxes is to be ascribed to their source spectra, which for
antideuterons is the square of the antiproton one.  This effect, added
to the different  weight of destruction cross sections, is visible
also in the lower energy tail of the spectrum.  The destruction of the
antideuteron nuclei on the ISM alters the flux by a factor of two,
while the antiproton one is modified by a mere 20--25\%.

\subsubsection{Dark Matter Halo profile uncertainty}

\begin{figure}[t] 
\centering 
\vspace{-42pt}
\includegraphics[width=1.1\columnwidth]{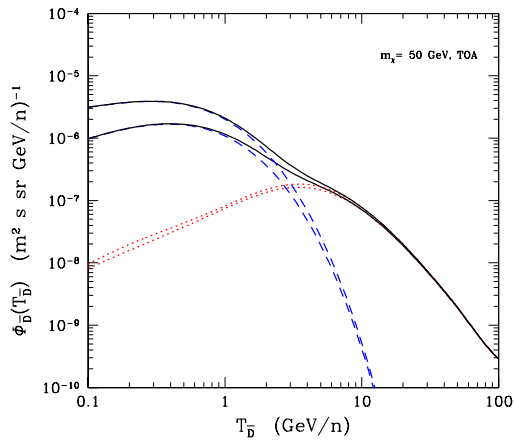} 
\vspace{-30pt} 
\caption{Interstellar and Top--Of--Atmosphere (TOA) antideuteron fluxes. The dashed (blue) line shows the
primary flux for  $m_\chi$=50 GeV and $\sigmav_0 = 2.3 \cdot 10^{-26}$ cm$^3$
s$^{-1}$, the (red) dotted line denotes the secondary component and the
(black) solid line stands for the total (signal+background) flux. Propagation
model is the median one in Table \ref{table:prop}.}
\label{m50_tot_is} 
\end{figure}

The effect of changing the  DM distribution function $\rho_{\rm DM}(r,z)$ on the 
\dbar\  flux is demonstrated in  Fig.~\ref{profiles}. We only modify the shape
of the density  distribution (as discussed in Sect.~\ref{sec:DM}), while
keeping frozen the local DM density to $\rho_\odot$=0.42 GeV/cm$^3$.  The DM
mass is $m_\chi$=50 GeV, but as explained in the previous Section, the source
term mostly factors out, so that these conclusions hold for any
neutralino mass.  We plot the ratio $(\phi_{\bar d} - \phi_{\bar
d}^{\rm ref})/\phi_{\bar d}^{\rm ref}$  where $\phi_{\bar d}^{\rm ref}$ is the
reference flux calculated with the cored isothermal profile and $\phi_{\bar
d}$ corresponds to the NFW, DMS--1.2 and N04 profiles (see
Sect.~\ref{sec:DM}).  The two classes of curves correspond to the maximal
(upper, solid) and median (lower, dotted) propagation parameters. 
The difference on the fluxes calculated with  the minimal set of propagation
parameters (not shown) is negligible. 

The increasing steepness of the profile in central regions of the Galaxy is 
responsible for an increasing of the \dbar\ flux which is more relevant for
higher  diffusive halos. In the case $L=15$ kpc, the  \dbar\ obtained with a
1.2 cuspy profile  \cite{2004MNRAS.353..624D} is a factor of 2 higher than the
cored one, while the NFW  \cite{1997ApJ...490..493N} halo gives fluxes 30-40\%
higher than the isothermal one depending on energy. The highest flux is 
obtained with the log-slope NFW-like profile of \citet{2004MNRAS.349.1039N},  
which predicts, among the considered DM profiles, the highest DM density in a
wide radial interval around the Solar System, although
it is flatter than the DMS--1.2 and NFW profiles in the central kpc of the Galaxy.  
The flux obtained with the
median parameters ($L=4$ kpc) is less significantly modified by  a change in
the halo profile. Indeed, a charged particle produced around the  galactic
center can more easily reach the Solar System when the magnetic diffusive halo
is larger and when it is more energetic.  

\section{Potential for Detection: results and discussion}
\label{discussion}

\begin{figure}[t] 
\centering 
\vspace{-42pt}
\includegraphics[width=1.1\columnwidth]{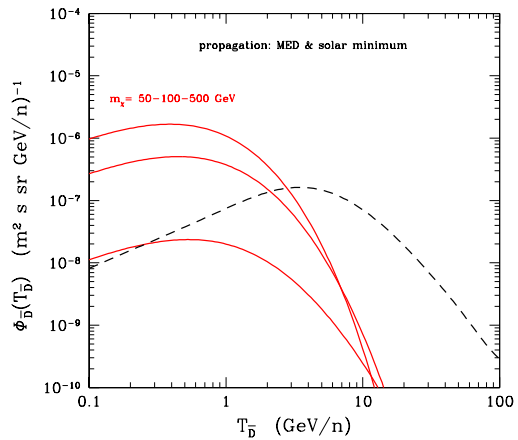} 
\vspace{-30pt} 
\caption{TOA fluxes for primary (solid lines) and secondary (dashed line) 
antideuterons for the median propagation parameters. From top to bottom, the solid lines
refer to WIMPs with mass $m_\chi$=50, 100, 500 GeV.}
\label{prim_sec} 
\end{figure}

\begin{figure}[t] 
\centering 
\vspace{-42pt}
\includegraphics[width=1.1\columnwidth]{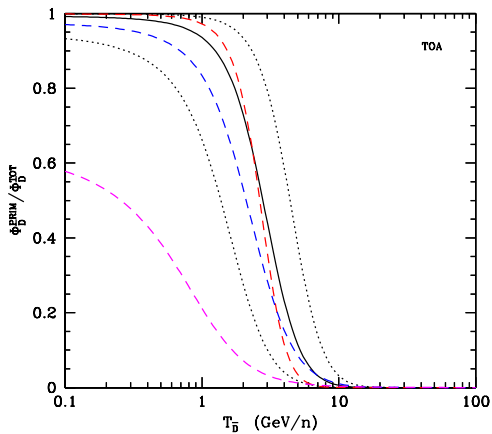} 
\vspace{-30pt} 
\caption{Ratio of the primary to total (signal+background) TOA antideuteron flux. 
Solid (black) curve refers to a WIMP mass of $m_\chi$=50 GeV and for the MED 
propagation parameters. Dotted (black) lines show the MAX (upper) and MIN (lower) cases. 
Dashed lines refer to the MED propagation parameters and different masses, which
are (from top to bottom): $m_\chi$=10, 100, 500 GeV (red, blue, magenta respectively).}
\label{ratio} 
\end{figure}

\begin{figure}[t] 
\centering 
\vspace{-42pt}
\includegraphics[width=1.1\columnwidth]{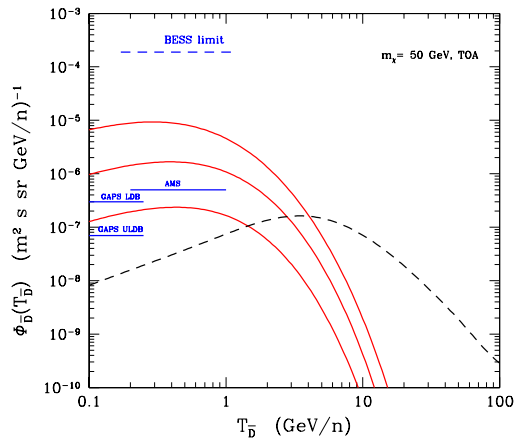} 
\vspace{-30pt} 
\caption{TOA primary (red solid lines) and secondary (black dashed line) 
antideuteron fluxes, modulated at solar minimum. The signal is derived for a 
$m_\chi$=50 GeV WIMP and for the three propagation models of Table \ref{table:prop}. 
The secondary flux
is shown for the median propagation model. The upper dashed horizontal line shows the
current BESS upper limit on the search for cosmic antideuterons. The 
three horizontal solid (blue) lines are the estimated sensitivities for (from
top to bottom): AMS--02 \cite{giovacchini}, GAPS on a long (LDB) and
ultra--long (ULDB) duration balloon flights \cite{2004NIMPB.214..122H,2006JCAP...01..007H,koglin_taup}.}
\label{futuro} 
\end{figure}

\begin{figure}[t] 
\centering 
\includegraphics[width=1.0\columnwidth]{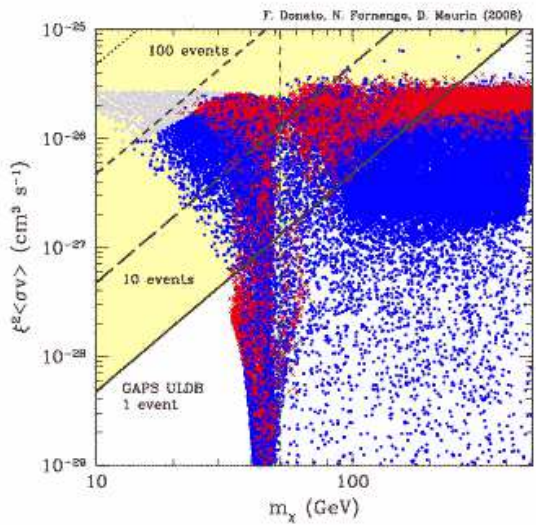} 
\vspace{-30pt} 
\caption{GAPS ULDB reach compared to predictions for neutralino dark matter in low--energy
supersymmetric models, shown in the plane effective annihilation cross section $\xisigmav_0$ vs. neutralino mass
$m_\chi$. The solid, long--dashed and short--dashed lines show our estimate for the capability of GAPS ULDB
of measuring 1, 10 and 100 events, respectively, for the median propagation model of Table \ref{table:prop}. The scatter plot reports the quantity $\xisigmav_0$ calculated
in a low--energy MSSM (for masses above the vertical [green] dashed line) and in non--universal gaugino
models which predict low--mass neutralinos \cite{Bottino:2008xc,Bottino:2007qg,Bottino:2005xy,Bottino:2004qi,Bottino:2003cz,Bottino:2003iu,Bottino:2002ry}. [Red] Crosses refer to cosmologically
dominant neutralinos, while [blue] dots stand for subdominant neutralinos. Grey point
are excluded by antiproton searches.}
\label{fig:susy-med} 
\end{figure}

\begin{figure}[t] 
\centering 
\includegraphics[width=1.0\columnwidth]{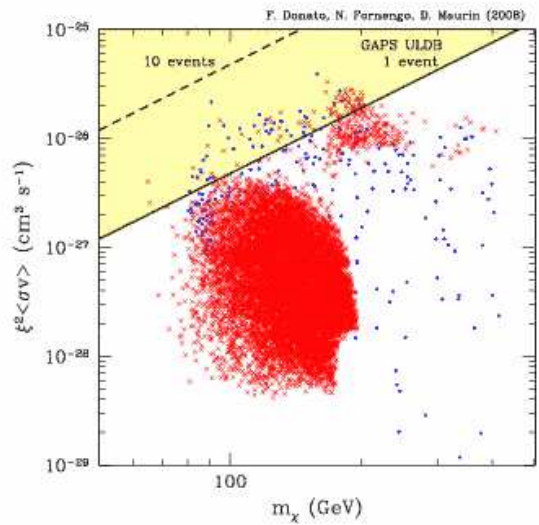} 
\vspace{-30pt} 
\caption{The same as in Fig.~\ref{fig:susy-med}, except that the supersymmetric predictions refer to a 
minimal SUGRA scheme.}
\label{fig:sugra-med} 
\end{figure}

\begin{figure*}[t] 
\centering 
\includegraphics[width=1.0\columnwidth]{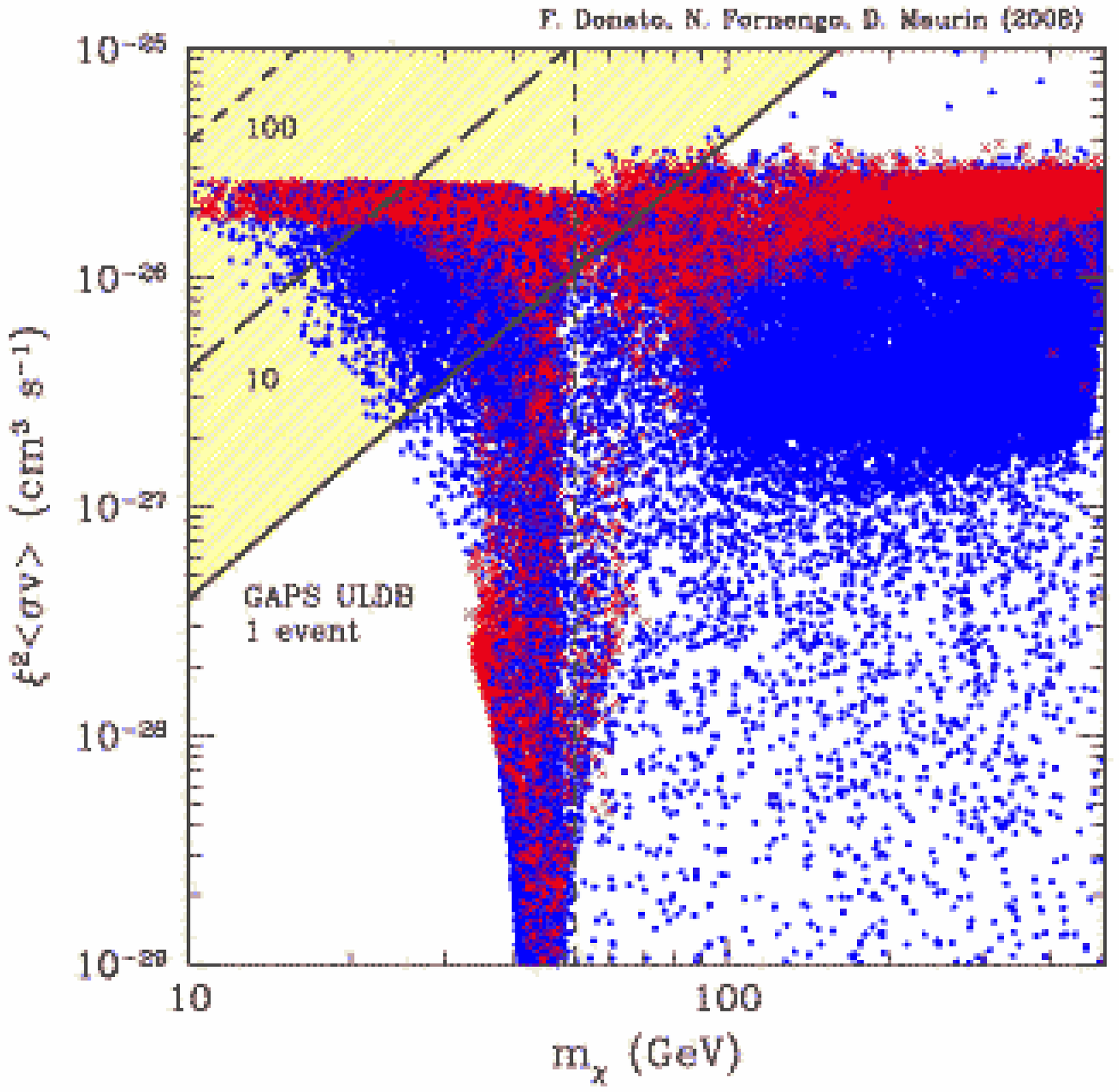} 
\includegraphics[width=1.0\columnwidth]{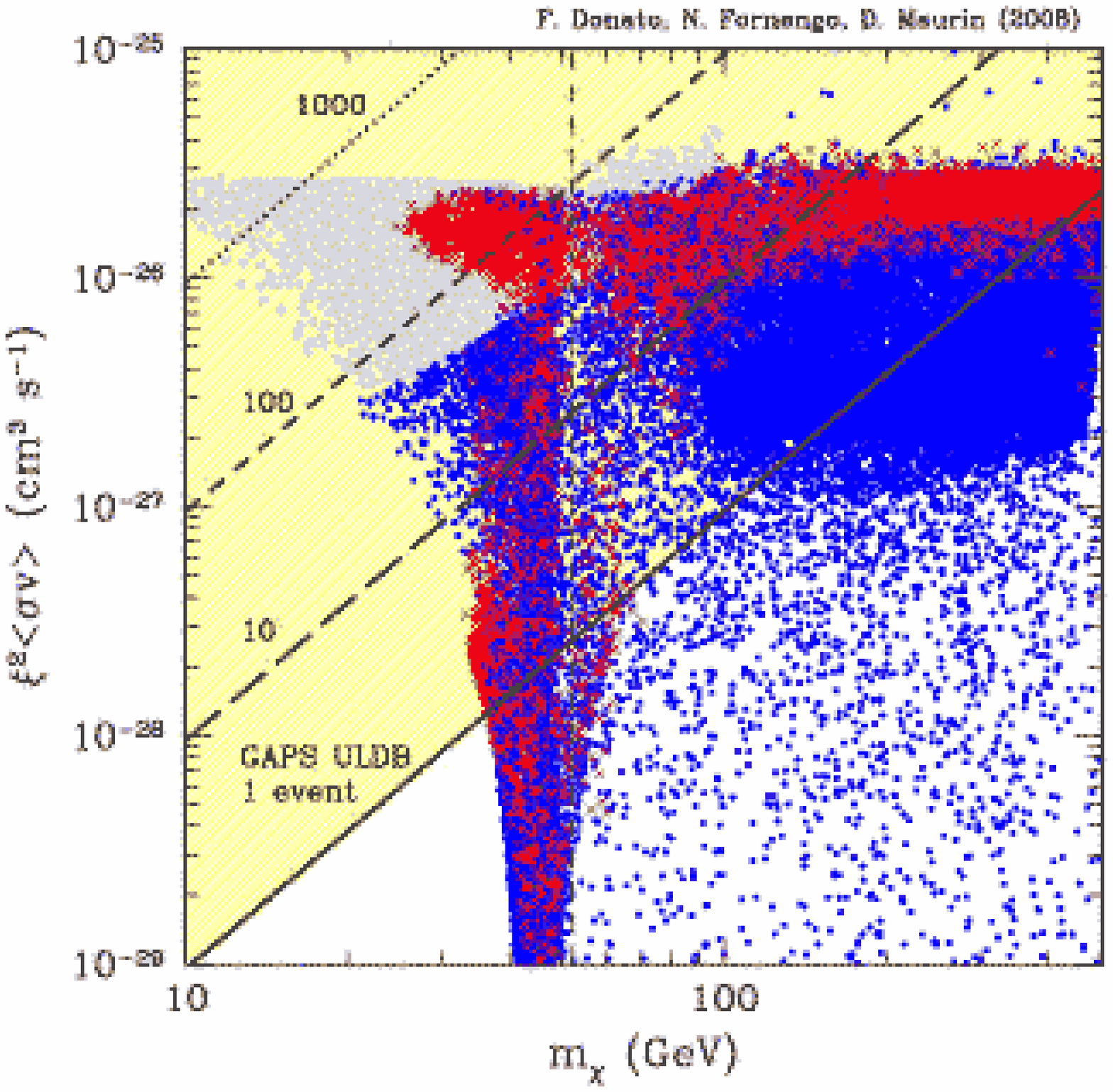} 
\vspace{-30pt} 
\caption{The same as in Fig.~\ref{fig:susy-med}, except that the astrophysical propagation parameters are those
which predict minimal (left panel) and maximal (right panel) antideuteron fluxes. In the right panel, grey point
are excluded by antiproton searches.}
\label{fig:susy-extreme} 
\end{figure*}

We now turn to the {determination of the} total flux we can expect from the standard astrophysical source
(see Sect.~\ref{secondary}) {added} to a possible contribution from DM annihilation. 
In Fig.~\ref{m50_tot_is}, we {show} the IS and TOA (solar 
minimum) secondary and primary \dbar\ fluxes, and their sum. The primary flux is for a WIMP with
mass $m_\chi=50$ GeV, annihilation cross section $\sigmav_0 = 2.3 \cdot 10^{-26}$ cm$^3$ s$^{-1}$, 
an isothermal profile and $\rho_\odot=0.42$ GeV/cm$^3$, as in the 
previous figures. 
The discrepancy between primary and 
secondary flux for $T_{\bar{d}}\lsim$ 2 GeV/n is striking.
A signal from DM annihilation as the one in our example would definitely {increase by a large amount} 
the number of 
expected antideuterons in the lowest energy bins with respect to the purely secondary 
flux. At 100 MeV/n the 
expected \dbar\ flux from a {cosmologically dominant DM particle of} 50 GeV mass
is two orders of magnitude larger than the 
secondary \dbar\ flux calculated within the same propagation model. 
{One has to remind that the primary flux scales as $m_\chi^2$: this means that, in the low
energy sector, the signal can overwhelm the background up to masses of the order of few hundreds of GeV.}
This figure demonstrates that the search for cosmic antideuterons is {definitely} one of 
the most powerful indirect detection means for the  DM annihilation in the halo of our Galaxy. 
The discrimination power between signal and background 
can be as high as few orders of magnitude. 
A major limit to this kind of experimental inspection 
may reside in the tiny level of the expected 
flux (about four orders of magnitude less abundant than antiprotons), which nevertheless is foreseen to become
experimentally accessible in the near future \cite{2004NIMPB.214..122H,2006JCAP...01..007H,koglin_taup,giovacchini}.

Figure~\ref{prim_sec} displays the TOA \dbar\ flux for the median
propagation parameters and at solar minimum. Together with the secondary
flux, we plot the primary one for three different WIMP masses: 50, 100, 500
GeV and for the same reference value of the annihilation cross section.
{As discussed above, lighter WIMPs would provide a striking signal,
and sensitivity is present for masses up to few hundreds of GeVs.}

The
discrimination power between primary and secondary \dbar\ flux may
also be deduced from  Fig.~\ref{ratio}.
The ratio of the primary to
total TOA \dbar\ flux is plotted as a function  of the kinetic energy
per nucleon, for the three representative propagation models and
different WIMP masses {(the annihilation cross section is again fixed at the
reference value)}. This ratio keeps higher than 0.7 for $T_{\bar{d}}<1$ GeV/n
except for $m_\chi=$500 GeV.  For propagation models with $L \gsim 4$
kpc -- which is a very reasonable expectation -- this ratio   is at
least 0.9 for masses below 100 GeV. Increasing the WIMP mass, we must
descend to lower  energies in order to maximize the
primary--to--secondary ratio.  However, for a  $m_\chi=$500 GeV WIMP
we still have a 50-60\% of DM contribution in the 0.1-0.5 GeV/n
range. Of course, the evaluation of the theoretical uncertainties
presented in this Paper must be  kept in mind while confronting to
real data.  Fig.~\ref{ratio} clearly states that the antideuteron
indirect DM detection technique is probably  the most powerful one for
low and intermediate WIMP--mass haloes.

We finally discuss in Fig.~\ref{futuro} a possible experimental 
short term scenario.  The secondary \dbar\ flux for the median
configuration of Table \ref{table:prop} is plotted  alongside the
primary flux from $m_\chi=$50 GeV, calculated for the maximal, median
and minimal  propagation scenarios.  The present BESS upper limit on
the (negative) antideuteron search \cite{2005PhRvL..95h1101F}  is at a
level of 2$\cdot 10^{-4}$ (m$^2$ s sr GeV/n)$^{-1}$. We also plot the
estimated sensitivities of the gaseous antiparticle spectrometer GAPS
on a long duration balloon flight (LDB) and  an ultra--long duration balloon
mission (ULDB) 
\cite{2004NIMPB.214..122H,2006JCAP...01..007H,koglin_taup}, 
and of AMS--02 for three years of data taking
\cite{giovacchini}.  The perspectives to explore a part of  the region
where DM annihilation are mostly expected (i.e. the low--energy tail) are  very promising.  If one of
these experiments will measure at least 1 antideuteron, it will be a
clear  signal of an exotic contribution to the cosmic antideuterons.
Note that for AMS, a sensitivity at the level of the one at low
  energy should be obtained beyond 2.3 GeV/n thanks
  to the RICH \cite{barao,giovacchini}. This higher energy region would be complementary to a low energy detection. 

Prospects for antideuteron searches for specific dark matter candidates are
shown in Figs.~\ref{fig:susy-med},~\ref{fig:sugra-med} and~\ref{fig:susy-extreme},
where the expected capabilities of GAPS ULDB are confronted with theoretical
predictions for neutralino dark matter in various supersymmetric schemes.
The results are shown in terms of the effective annihilation cross section
$\xisigmav_{0}$ as a function of the neutralino mass. We recall that
DM annihilation signals are proportional to the square of the relic--particle 
density profile $\rho_{\chi}$ (which is in turn proportional to the total DM density through the
rescaling factor $\xi$ defined
in Sect.~\ref{primary}), to the low--energy thermally--averaged annihilation cross
section $\sigmav_{0}$ and inversely proportional to the square of the WIMP mass
$m_{\chi}$. It is therefore convenient to show our prediction in the
plane $\xisigmav_{0}$ -- $m_{\chi}$. The solid, long--dashed and short--dashed
lines in Fig.~\ref{fig:susy-med} denote our estimate for GAPS ULDB to measure 1, 10 and
100 events, respectively. The number of events has been obtained by using an
acceptance $\times$ time for the GAPS detector in its whole energy range of 
$1.33 \times 10^7 {\rm m}^2 {\rm sr~s~GeV/n}$ \cite{hailey}, for which 
no contribution to the antideuteron signal is expected from the secondary component. The
quoted number of events therefore refers uniquely to the signal component. The transport parameters used
in the determination of the antideuteron fluxes are
those which provide the median flux and a minimum of activity for the Sun has been considered for the
solar modulation.
For definiteness, the event--lines have been calculated for
the case of a pure $b\bar b$ DM annihilation final state. Fig.~\ref{ann_state} shows
that no major differences are present when different final states are considered: we therefore consider
the lines of Fig.~\ref{fig:susy-med} as a good estimate of the GAPS ULDB sensitivity, to be confronted
with theoretical predictions for specific candidates.

The scatter plot
shows the distribution of points obtained in two supersymmetric models: for masses
larger than about 50 GeV, the points refer to a low--energy Minimal Supersymmetric Standard
Model (MSSM), where all the mass parameters are defined at the electroweak scale \cite{Bottino:2000jx,2004PhRvD..69f3501D}. For masses
lighter than 50 GeV, the scatter plots refers to the light--mass neutralinos in
the models of Refs. \cite{Bottino:2008xc,Bottino:2007qg,Bottino:2005xy,Bottino:2004qi,Bottino:2003cz,Bottino:2003iu,Bottino:2002ry} where the gaugino universality condition is violated.
The (red) crosses denote configurations for which the relic neutralino is a dominant dark matter
component (i.e. its calculated relic abundance falls in the WMAP range \cite{Hinshaw:2008kr,Komatsu:2008hk,Dunkley:2008ie} for the
CDM content of the Universe), while the (blue) points refer to subdominant neutralinos (for these points 
$\xi=\Omega_{\chi} h^{2}/(\Omega_{\rm CDM}h^{2})_{\rm WMAP}$ applies). 
In the case of the standard MSSM, our prediction is that up to about a 20 events could be detected, for
masses between 50 and  100 GeV. In the case
of the low--mass neutralino models, a large number of events (up to a few hundreds of events
for 10 GeV neutralinos)
are expected for most of the relevant
supersymmetric parameter space. In this situation, where the antideuteron fluxes are large, 
also antiproton production is expected to be sizeable. We therefore
need to check that we do not produce an excess of exotic antiprotons which goes in conflict with
current measurements on this antimatter channel. We employ here a 2--$\sigma$ upper bound on the
admissible excess of antiprotons over the background in the lowest energy bin of BESS \cite{Orito:1999re,
Maeno:2000qx} and AMS \cite{Aguilar:2002ad} of 
$0.68 \times 10^{-2} {\rm m}^{-2} {\rm sr}^{-2} {\rm s}^{-1} {\rm GeV}^{-1}$. These configurations,
which exceed the antiproton bound \cite{Bottino:2005xy}, are shown by the gray points. We see that,
once the $\bar p$ bound is included, up to about 100 \dbar\  events are expected for neutralinos
of 15--20 GeV.A detector like GAPS ULDB will therefore
have a strong capability to access this class of supersymmetric models, which are currently 
only partially probed by antiprotons searches \cite{Bottino:2005xy} and direct detection \cite{Bottino:2005qj}, depending on the actual galactic parameters, like profile and local density. 
In the case of the standard MSSM, GAPS ULDB sensitivity allows to probe neutralino masses up to 300 GeV.

The case of supersymmetric models endowed with Supergravity (SUGRA) boundary conditions  \cite{Berezinsky:1995cj} are showed in Fig.~\ref{fig:sugra-med} for the same astrophysical models of
Fig.~\ref{fig:susy-med}, together with our GAPS ULDB expectations. In the case of minimal SUGRA we predict up to a few events, for some
specific supersymmetric configurations, although the bulk of the parameter space requires larger
sensitivities to be probed.

The cases of the minimal and maximal set of transport parameters is shown in Fig.~\ref{fig:susy-extreme},
again for the MSSM and gaugino non--universal models. In the case of the ``min'' set of parameters, 
light neutralinos may produce up to 70 events, while the in the standard MSSM at most a few events
are predicted. For the ``max'' case we again are confronted with
the fact that
an excess of exotic antiprotons in conflict with
current $\bar p$ measurements plays a role. Once this bound is properly taken into account, we obtain that
100--300 antideuterons are expected in the low/mid--mass range for neutralinos. Also in the standard MSSM model,
up to 40 events are foreseen. In this case, the sensitivity of GAPS ULDB extends up to neutralino
masses of the order of 500 GeV.

In conclusion, antideuteron searches offer very good opportunity of detecting a signal, for a variety of supersymmetric schemes. Especially low mass neutralinos in gaugino non--universal models \cite{Bottino:2008xc,Bottino:2007qg,Bottino:2005xy,Bottino:2004qi,Bottino:2003cz,Bottino:2003iu,Bottino:2002ry} provide expectations for a large number of events in a detector like GAPS ULDB, with good opportunities to
clearly disentangle a signal.

\section{Conclusions}
\label{sec:conclusions}

In this Paper we present a novel and updated calculation of both the secondary
and primary antideuteron fluxes, with special attention to the determination
of the uncertainties of nuclear and astrophysical origin which affect the
theoretical predictions of the \dbar\ flux. The galactic environment is
treated in a two--zone diffusion model, the same that successfully reproduces
cosmic--ray nuclear data, like e.g. the boron--to--carbon ratio
\cite{2001ApJ...555..585M,2002A&A...394.1039M}, and that is able to predict
the observed antiproton flux \cite{2004PhRvD..69f3501D}. We therefore refine
our calculation of Ref.~\cite{2000PhRvD..62d3003D}, where the antideuteron
signal as a promising tool for dark matter searches was originally proposed.

We review the nuclear and astrophysical uncertainties and provide an up to
date secondary ({\em i.e.} background) antideuteron flux.  Propagation
uncertainties for the secondary component range from 40-50 \% {around }the
average {flux} at energies below {1--2 GeV/n} down to $\sim 15$ \% at 10
GeV/n.  Nuclear uncertainties are largely dominant: a generous
and conservative factor of 10 at very low
energies, which reduces to a factor of four at 100 GeV/n.

The primary contribution has been calculated for generic WIMPs 
annihilating in the galactic halo, for the different production channels
through which the antideuteron signal may be produced by DM annihilation.
As for the antiproton \cite{2004PhRvD..69f3501D} and positron signals \cite{Delahaye:2007fr}, we 
obtain that the transport--related processes induce the largest source of uncertainty in the antideuteron
flux: it ranges from a factor of 10 upward and downward with respect to the median, best--fit,
prediction, in the low energy range.

Our theoretical predictions have then been confronted with the expected sensitivities of future
detectors, specifically GAPS 
\cite{2006JCAP...01..007H,2004NIMPB.214..122H,2007NuPhS.173...75K}
and AMS \cite{Aguilar:2002ad,2007arXiv0710.0993A,giovacchini}. We have considered neutralinos
as dark matter candidates and discussed three specific supersymmetric scenarios: a low--energy
MSSM scheme, a gaugino non--universal supersymmetric model and a minimal SUGRA framework.
For these classes of models we have analyzed the potentiality of
the GAPS detector in a ultra long duration balloon flight and found that this detector will have
the capabilities to detect up to a few hundred events for low--mass neutralinos (in the mass
range from 10 to 50 GeV) in the gaugino non--universal models, and up to tens of events for 50-100
GeV neutralinos in standard low--energy MSSM. The sensitivity of GAPS ULDB on the
neutralino mass extends up to 300-500 GeV in the low--energy MSSM.

In conclusion, antideuterons offer an exciting target for indirect dark
matter detection for low and intermediate WIMP masses and future experiments
will have a unique opportunity to clearly identify a signal.

\acknowledgments
D.M thanks R.~Duperray for sharing the \dbar\ cross section files,
and R.~Taillet and P.~Salati for their help during several debugging
sessions of the propagation code.
We thank P. Salati for useful discussions and comments in the
early stage of this study. We finally thank C. Hailey for
providing us with informations about the GAPS detector and F. Barao
for the AMS detector.
N.F. and F.D. acknowledge research grants funded jointly by the Italian Ministero
dell'Istruzione, dell'Universit\`a e della Ricerca (MIUR), by the
University of Torino and by the Istituto Nazionale di Fisica Nucleare
(INFN) within the {\sl Astroparticle Physics Project}.

\appendix
\section{Cross sections}
\label{App:xsec}

\begin{figure}[t] 
\centering 
\includegraphics[width=1.\columnwidth]{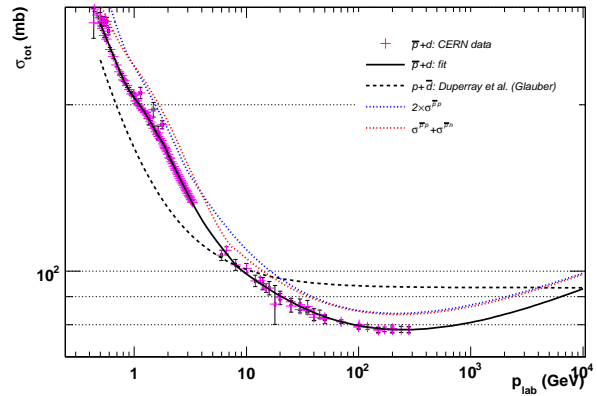} 
\caption{Modeling of the total (elastic+inelastic) cross section
$\sigma^{H+\bar{d}}_{tot}$. All curves are displayed as a function
of the momentum. The black dashed line corresponds to the Glauber
approximation used in \citet{2005PhRvD..71h3013D}. Crosses refer to data
for the charge conjugate reaction $\bar{p}+d$ \citep{PDBook}. The black solid
line is a crude fit of the latter reaction when data are available (the
high energy regime comes from Ref.~\citep{PDBook}). The two dotted curves
illustrate a tentative estimate of $\sigma^{\bar{d}H}$ as $2\sigma^{\bar{p}p}$
(blue) and $\sigma^{\bar{p}p}+\sigma^{\bar{p}n}$ (red), where we have
modelled the latter cross sections by fits on the available
data \citep{PDBook}.} 
\label{fig:tot} 
\end{figure} 

\begin{figure}[t] 
\centering 
\includegraphics[width=1.\columnwidth]{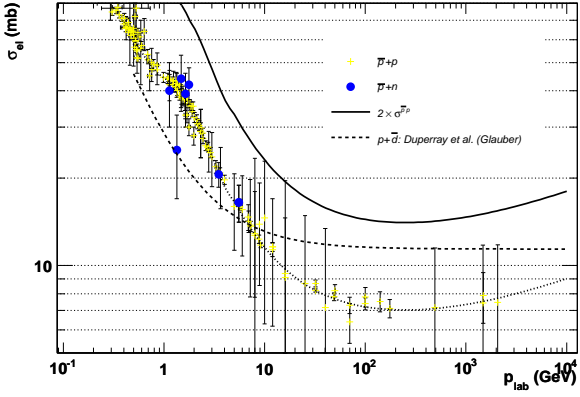} 
\caption{Elastic cross sections as a function of the incident particle
momentum. Yellow crosses and blue circles are CERN data \citep{PDBook}
for $\bar{p}+p$ (the dotted line is a fit to these data) and $\bar{p}+n$
reactions. The elastic cross section $\sigma^{p\bar{d}}_{el}$ is
modelled using $2\sigma^{\bar{p}p}_{el}$ (solid black line).
The dash-dotted grey line corresponds to the Glauber cross section
used in \citet{2005PhRvD..71h3013D}. See text for comments.}
\label{fig:el} 
\end{figure}

\begin{figure}[t] 
\centering 
\includegraphics[width=1.\columnwidth]{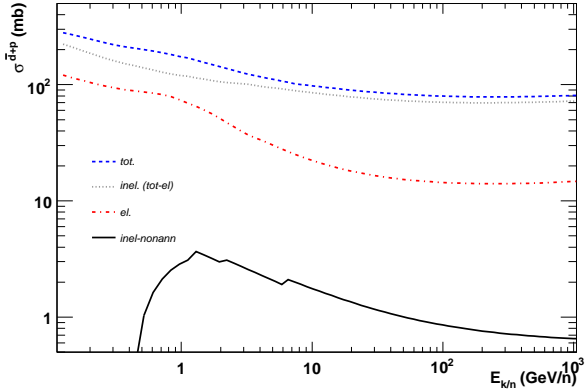} 
\caption{From top to bottom: total, inelastic, elastic and inelastic non-annihilating 
cross sections for $\bar{d}+H$ as a function of the \dbar\ kinetic energy per nucleon.}
\label{fig:all} 
\end{figure}

We review here the cross sections (total, elastic, inelastic
and non-annihilating) used in the calculation for primary
or secondary \dbar. Their detailed and thorough study, including
the production, has been presented in \citet{2005PhRvD..71h3013D}.
We do not repeat their arguments, but rather {recall} the main
characteristics of these cross sections, providing alternative
formulations, minor corrections or further checks when possible.
However, none of these ``updated" cross sections lead to
significant changes in the calculation of the \dbar\
flux, compared to that calculated with the cross sections
presented in Ref.~\citep{2005PhRvD..71h3013D}. The main source
of uncertainty remains the production cross section discussed
in Sect.~\ref{sec:production}.

\subsection{Total and elastic cross section}
In \citet{2005PhRvD..71h3013D}, the Glauber approximation was used.
No data exist for the total cross section of the process $\bar{d}+H$,
but there are measurements for the charge conjugate reaction $d+\bar{p}$.
Figure \ref{fig:tot} shows various modeling of
$\sigma^{\bar{d}p}_{\rm tot}$. Along with the charge conjugate
measurements and the associated fit function, we also plot
the Glauber cross section \citep{2005PhRvD..71h3013D} and two modelings using
$\bar{p}p$ and $\bar{p}n$ data. In this Paper,
we assume $\sigma^{\bar{d}p}_{\rm tot}=\sigma^{d\bar{p}}_{\rm tot}$,
so that the total cross section is given by the black solid line
of Fig.~\ref{fig:tot}.
Note that the combination $2\sigma^{\bar{p}p}$ and 
$\sigma^{\bar{p}p}+\sigma^{\bar{p}n}$ are very close
to $\sigma^{d\bar{p}}$ data, overshooting it by a mere 10\%.

Concerning the elastic cross section, the situation is worse
{since} no data exist even for the charge conjugate reaction.
However, as seen in Fig.~\ref{fig:tot},
the total cross section $\sigma^{d\bar{p}}$ is well approximated
by $2\sigma^{\bar{p}p}$ or $\sigma^{\bar{p}p}+\sigma^{\bar{p}n}$,
so {this} should be the also case for the elastic cross section.
{Since} data for $\bar{p}n$ are scarce (Fig.~\ref{fig:el}, blue circles)
as compared to those for $\bar{p}p$ (yellow crosses),
we choose to approximate $\sigma^{p+\bar{d}}_{\rm el}\approx
2\sigma^{\bar{p}p}_{\rm el}$. This is shown on Fig.~\ref{fig:el}
as a black solid line, which is a factor of two larger
than the Glauber description (black dashed line) used
in Ref.~\cite{2005PhRvD..71h3013D}.

\subsection{Inelastic, total non-annihilating and differential redistribution cross sections }

\begin{figure}[t] 
\centering 
\includegraphics[width=1.\columnwidth]{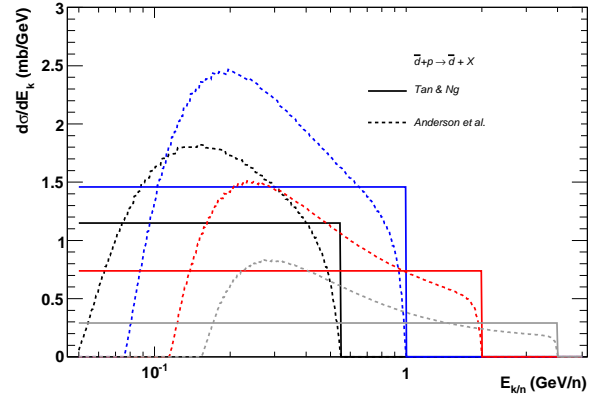} 
\caption{Differential redistribution cross section $\bar{d}+p\rightarrow \bar{d}+X$
as a function of the \dbar\ kinetic energy per nucleon. Solid curves are
for the limiting fragmentation hypothesis as used, e.g., in
Tan \& Ng \citep{1983JPhG....9..227T} and
dashed curves are for the Anderson et al. parameterization \citep{1967PhRvL..19..198A}.
The four sets of curves correspond to four incident \dbar\ energy:
0.55~GeV/n (black), 1~GeV/n (blue), 2~GeV/n (red) and 4~GeV/n (grey).}
\label{fig:ter} 
\end{figure}

The inelastic cross section is simply obtained from:
\begin{equation}
\sigma_{\rm inel}=\sigma_{\rm tot}-\sigma_{\rm el}\,,
\end{equation}
and is itself further expressed as:
\begin{equation}
\sigma_{\rm inel} =
\sigma_{\rm inel}^{\rm non-ann} + \sigma_{\rm inel}^{\rm ann}\,.
\end{equation}
The non-annihilating (non-ann) part corresponds to antideuterons
that interact inelastically with H, but survive the collision,
loosing a fraction of their initial energy.

In Ref.~\citep{2000PhRvD..62d3003D}, the tertiary contribution was neglected,
arguing that \dbar\ (or symmetrically $d$) incident on a nucleon or
on a nuclear target should have a small non-annihilating cross section
owing of the small \dbar\ ($d$) nuclear binding energy.
As discussed in Ref.~\citep{2005PhRvD..71h3013D}, this intuitive argument
can be invalidated both on empirical evidence and on formal 
grounds (see discussion and references therein). We stick to the
empirical approach by using the cross section given in Ref.~\citep{2005PhRvD..71h3013D}. 
The total inelastic non-annihilating cross section $\sigma^{\bar{d}+p}_{\rm inel, non-ann}$
(NAR in Ref.~\citep{2005PhRvD..71h3013D}) is obtained by summing up the
$\bar{p}+d\rightarrow (n\pi)\bar{p}d$ cross sections which are experimentally available.
As seen in Fig.~\ref{fig:all} (black solid line), the cross section peaks
at $\approx 4$~mb. As no attempt was made to evaluate the (expectedly small)
contributions of the channels not known experimentally, the overall evaluation
is a lower bound \citep{2005PhRvD..71h3013D}.
In the same Fig.~\ref{fig:all}, for reference purposes, we show all the cross
sections used in this Paper.

The last issue is the energy redistribution of the non-annihilated \dbar.
In most of the papers dealing with antinuclei
\citep[e.g.][]{2000PhRvD..62d3003D,2001ApJ...563..172D,1999ApJ...526..215B}, the energy distribution
of the surviving particles is based on the limiting fragmentation hypothesis
\citep[see references in][]{2005PhRvD..71h3013D}:
\begin{equation}
	\frac{d\sigma^{\bar{d}H\rightarrow{\bar{d}X}}}{dE_k}
	   (E_k' \rightarrow E_k) = 
		\frac{\sigma^{\bar{d}H}_{\rm inel, non-ann}}{E_k'}\;.
\end{equation}
However, experimentally, the $pp\rightarrow pX$ differential cross section
was shown to be largely independent of the longitudinal momentum $p_l^*$ of
the produced particles in the center of mass and can be written in
the laboratory frame as \citep{1967PhRvL..19..198A}:
\begin{eqnarray}
&&\!\!\!\!\!\!\!\!\!\!\!\!\!\!\!\!\frac{d^{2}\sigma(pp\rightarrow pX)}{dpd\Omega}=\nonumber \\
&&\frac{p^2}{2\pi
p_t} \frac{\gamma(E-\beta p\cos\theta)}{E} \, 610 p_t^{2}\,
\exp{\left[{-\frac{p_t}{0.166}} \label{eq:5-25}\right]}
\end{eqnarray}
where $\gamma$ and $\beta$ are the usual Lorentz factor and particle velocity, and $p_t$
the transverse momentum of the particle.
Following \citet{2005PhRvD..71h3013D}, the same functional form is used 
for \dbar\ and we define:
\begin{equation}
\frac{d\sigma^{\rm Anderson}}{dE_k} \equiv \int_{\theta}^{2\pi}
\frac{d^{2}\sigma(pp\rightarrow pX)}{dpd\Omega} \sin\theta d\theta.
\end{equation}
Note that an incorrect normalization was applied in Ref.~\citep{2005PhRvD..71h3013D}
for the final cross section. The differential cross section for the tertiary term reads:
\begin{eqnarray}
	\!\!\!\!\frac{d\sigma^{\bar{d}H\rightarrow{\bar{d}X}}}{dE_k}
	   \!\!\!\!\!\!& &(E_k' \rightarrow E_k) = \sigma^{\bar{d}H}_{\rm inel, non-ann} \nonumber\\
		 \!\!\!\!\times\!\!\!\!&&
		  \left[ \frac{d\sigma^{\rm Anderson}}{dE_k} \right] \cdot
			\left[ \int_0^{E_k'} \frac{d\sigma^{\rm Anderson}}{dE_k''} dE_k''\right]^{-1}.	 
\end{eqnarray}
Note, however, that this correction does not change significantly the tertiary
spectrum and conclusions given in Ref.~\citep{2005PhRvD..71h3013D}. The inelastic
scattering spectra in the laboratory for 0.55 (black), 1 (blue), 2 (red)
and 4~GeV/n (grey) \dbar\ are shown in Fig.~\ref{fig:ter}. For each of these energies,
the Anderson et al. \citep{1967PhRvL..19..198A} scheme more 
efficiently redistributes \dbar\ at very
low energy (a few hundreds of MeV), as compared with the other one.

Finally, we have not yet discussed  the $\bar{d}+$He cross sections. Here, for
the total, elastic, non-annihilating cross section, we simply use the naive geometrical
factor $A^{2/3}$ to account for interaction on He gas in the ISM:
\begin{equation}
	\sigma^{\bar{d}+He}\approx 2.52\, \sigma^{\bar{d}+H}.
\end{equation}
The same factor is also assumed for the differential tertiary cross section.

\bibliography{dfm}

\end{document}